# The Atomic and Electronic structure of 0° and 60° grain boundaries in MoS$_2$

Terunobu Nakanishi[1], Shoji Yoshida[2], Kota Murase[2], Osamu Takeuchi[2], Takashi Taniguchi[3], Kenji Watanabe[3], Hidemi Shigekawa[2], Yu Kobayashi[4], Yasumitsu Miyata[4], Hisanori Shinohara[1], and Ryo Kitaura[1,*]

1 Department of Chemistry, Nagoya University, Nagoya 464-8602, Japan
2 National Institute for Materials Science, 1-1 Namiki, Tsukuba 305-0044, Japan
3 Faculty of Pure and Applied Sciences, University of Tsukuba, Tsukuba 305-8571, Japan
4 Department of Physics, Tokyo Metropolitan University, Hachioji, Tokyo 192-0397, Japan

*Corresponding Author: R. Kitaura
Tel: +81-52-789-2482, Fax: +81-52-747-6442,
E-mail: r.kitaura@nagoya-u.jp

ABSTRACT

We have investigated atomic and electronic structure of grain boundaries in monolayer MoS$_2$, where relative angles between two different grains are 0 and 60 degree. The grain boundaries with specific relative angle have been formed with chemical vapor deposition growth on graphite and hexagonal boron nitride flakes; van der Waals interlayer interaction between MoS$_2$ and the flakes restricts the relative angle between two different grains of MoS$_2$. Through scanning tunneling microscopy and spectroscopy measurements, we have found that the perfectly stitched structure between two different grains of MoS$_2$ was realized in the case of the 0 degree grain boundary. We also found that even with the perfectly stitched structure, valence band maximum and conduction band minimum shows significant blue shift, which probably arise from lattice strain at the boundary.

INTRODUCTION

A post-graphene material, transition metal dichalcogenide (TMD), has recently attracted a great deal of attention. TMDs have a long research history, but research on properties of monolayer TMDs, three-atom-thick atomic layers, has only recently been started.[1-3] One of the most distinct in TMDs from graphene is that TMDs can have sizable bandgap (~2 eV), leading to electronic and optoelectronic applications of TMD atomic layers.[4] In fact, various TMD-based devices, including high-performance FET devices, light-emitting transistors, and photodetectors, have actually been demonstrated.[5-7] In conjunction with the flexibility arising from the ultrathin structure, flexible electronic and optoelectronic devices can also be made.[8,9] In addition, monolayer TMDs in 2H form can have valley-degree-of-freedom, which may lead to future novel electronic devices based on valleytronics.[10,11]

For future applications of TMDs for electronic and optoelectronic devices, wafer-scale monolayer TMDs grown by chemical vapor deposition (CVD) are indispensable.[12,13] Top-down approaches, such as mechanical exfoliation, are not compatible with wafer-scale monolayer TMDs, and a bottom-up approach is required for that purpose.[14] Crystal growth by CVD is a bottom-up approach to obtain thin films, having been successfully applied to grow various atomic layers, such as graphene, hexagonal boron nitrides (hBN), and TMDs.[15-20] In typical CVD growth of TMDs, solid sources such as metal oxides and elemental sulfur are used, and monolayer TMDs film with a lateral size of millimeters have been reported.[19,21] Recently, the growth of TMDs by metal-organic CVD (MOCVD) with volatile liquid sources has been successfully demonstrated, and MOCVD is a promising method to realize wafer-scale TMDs that are compatible with device applications.[22-24]

In CVD-grown large-area TMDs, grain boundaries (GBs) are inevitably formed, which can significantly alter the electronic and optical properties of TMDs.[25-28] During the CVD growth of TMDs, nuclei form at the beginning of the CVD process, growing to form a large-area continuous sheet of TMD with GBs. Because the orientation of nuclei is normally random, a wide variety of GB structures can be formed. For example, a GB with 7-5 and 8-4-4 membered rings forms in a CVD-grown $MoS_2$, where midgap boundary states appear.[25,29-31] The existence of GB-induced midgap states significantly affects electronic transport across the boundary, leading to reduction of carrier mobility via additional carrier scattering at the GB.[25] Therefore, control of GB structure by controlling grain orientation and understanding the boundary-oriented electronic structure provide a basis for the realization of future TMD-based devices.

In this work, we have focused on orientation-limited growth of a TMD and investigation of localized boundary states using scanning tunneling microscopy (STM) and scanning tunneling spectroscopy (STS); the STS is a powerful tool to investigate domain boundaries[32,33]. The key for the successful control of crystal orientation in CVD growth of TMDs is the interaction between TMDs and the substrates used in CVD processes. In conventional CVD growth of TMDs, SiO$_2$/Si substrates with amorphous surfaces are used, leading to random crystal orientations of grown TMDs. In contrast, substrates with crystalline structures can limit the crystal orientation of grown TMDs through TMD-substrate interactions.[20,29,34,35] For the control of crystal orientation, we used hBN and graphite as substrates for CVD growth of TMDs. The atomically flat surfaces with three-fold (hBN) and six-fold rotation (graphite) symmetries successfully limited crystal orientations of grown TMD flakes; only two different orientations were observed. GBs between MoS$_2$ flakes with different orientations (relative angle of 60°) shows boundary states localized at specific location near the Fermi level. On the other hand, a GB between MoS$_2$ flakes with the same orientation shows a perfectly-stitched structure without any defects in scanned areas in STM images. We also found that both the conduction band minimum (CBM) and the valence band maximum (VBM) shift to the higher energy side at the GB even with a perfectly-stitched structure. This means that the GB state does not arise from defects but from strain at the GB, and strain formed at the growth process cannot be released even with the low friction coefficient between MoS$_2$ and graphite.

Methodology

We grew monolayer MoS$_2$ on hBN and graphite (Kish graphite, type C, Covalent Materials) flakes exfoliated on quartz substrates with a multi-furnace CVD apparatus. We prepared hBN and graphite flakes by the mechanical exfoliation method with adhesive tape (Scotch tape, 3M). As precursors for growth of MoS$_2$, we used molybdenum trioxide (Sigma-Aldrich, 99.5% purity) and sulfur powder (Sigma-Aldrich, 99.98% purity). Furnace temperatures at the locations where molybdenum trioxide and elemental sulfur were placed were set to 1029 K and 473 K, respectively, and the growth of MoS$_2$ was carried out at 1373 K for 20 min under Ar flow with a flow rate of 200 sccm. Atomic force microscope (AFM) observations were performed by the Veeco AFM system (Dimension 3100SPM, Nanoscope IV) operated at a scanning rate of 1 Hz. We measured photoluminescence (PL) spectra by a microspectroscopy system with a confocal microscope (Jobin Yvon HR-800, Horiba) with an excitation laser wavelength of 488 nm. For PL imaging, an LED light source (Mightex GCS-6500-15) was used to illuminate samples, and PL intensity ($\lambda$ > 600 nm) was imaged with CCD

(Princeton Instruments PIXIS-1024BR-eXelon). We formed electrical contacts to samples for STM/STS measurements by deposition of gold though a shadow mask or patterning conductive silver paste. After making the electrical contact, samples were introduced to an ultrahigh vacuum (UHV) environment and degassed at 473 K. The STM/STS measurements were conducted using a scanning tunneling microscope (Omicron LT–STM) in constant current mode operated at 90 K with an electrochemically etched W tip coated with PtIr (UNISOKU Co., Ltd.). A numerical derivative was used to acquire dI/dV curves, and WSxM software was used to process the STM images.[36]

Results and Discussion

PL imaging and spectroscopy have clearly shown that the quality of the present samples is high. Figure S1 shows a typical PL image of $MoS_2$/hBN and typical PL spectrum of $MoS_2$/hBN and $MoS_2$/graphite. As clearly seen, the PL image shows bright and uniform contrast, which clearly demonstrates high quality of samples we use. The observed FWHM values of PL spectra are 35 ~ 45 meV, which are much smaller than those of samples exfoliated onto $SiO_2$ substrates[37,38]. These PL spectra clearly demonstrate that quality of our sample is high.

The crystal orientations of $MoS_2$ grown on hBN and graphite are limited to two orientations due to the van der Waals interactions between $MoS_2$ and hBN. Figure 1(a) shows an AFM image of monolayer $MoS_2$ crystals grown on a hBN flake. As clearly seen, all crystals possess a hexagonal shape with long and short facets, and their orientations are limited to only two different ones, where 60° rotation of one orientation matches the other orientation. The observed long and short facets in the crystals correspond to chalcogen and metal zigzag edges; the relationship between crystal shape and crystallographic orientation was investigated with transmission electron microscopy and electron diffraction (additional information is given in Figure S2). Figure 1(b) shows structural models of hexagonal $MoS_2$ flakes with the two different orientations. The limited orientations of $MoS_2$ flakes are also observed in $MoS_2$ flakes grown on graphite substrates. This clearly demonstrates that the orientation-dependent potential arising from crystalline substrates is crucial to limiting the crystal orientation of grown $MoS_2$.

Because the crystal orientation of MoS$_2$ on hBN is limited, the resulting structure of the GBs should also be limited: GBs between grains with the same orientation (GB-0°) and grains with 60° mutual orientation (GB-60°). To investigate if defects exist at such GBs, we investigated the reactivity for an oxidation reaction. MoS$_2$ flakes with GBs were heated at 573 K under a flow of dry air. It has been shown that defects are sensitive to oxidation and reactions under the conditions above lead to the formation of oxides. Because oxidation from MoS$_2$ to the corresponding oxides heightens the pristine structure, position-sensitive detection of oxidation of MoS$_2$ can easily be done through AFM height images. Figures 1(c) and (e) ((d) and (f)) are AFM images of pristine (oxidized) MoS$_2$ flakes that have GB-0° and GB-60°, respectively. As clearly seen in Figs. 1(d) and (f), oxidation at Mo zigzag edges (shorter edges) is faster than that at S zigzag edges (longer edges).[39,40] We also found that GB-60° is oxidized as edges are oxidized, whereas GB-0° essentially retains its pristine structure. This means that GB-0° does not have defects that are sensitive to oxidation reactions, indicating that, unlike GB-60°, GB-0° has a well-stitched structure.

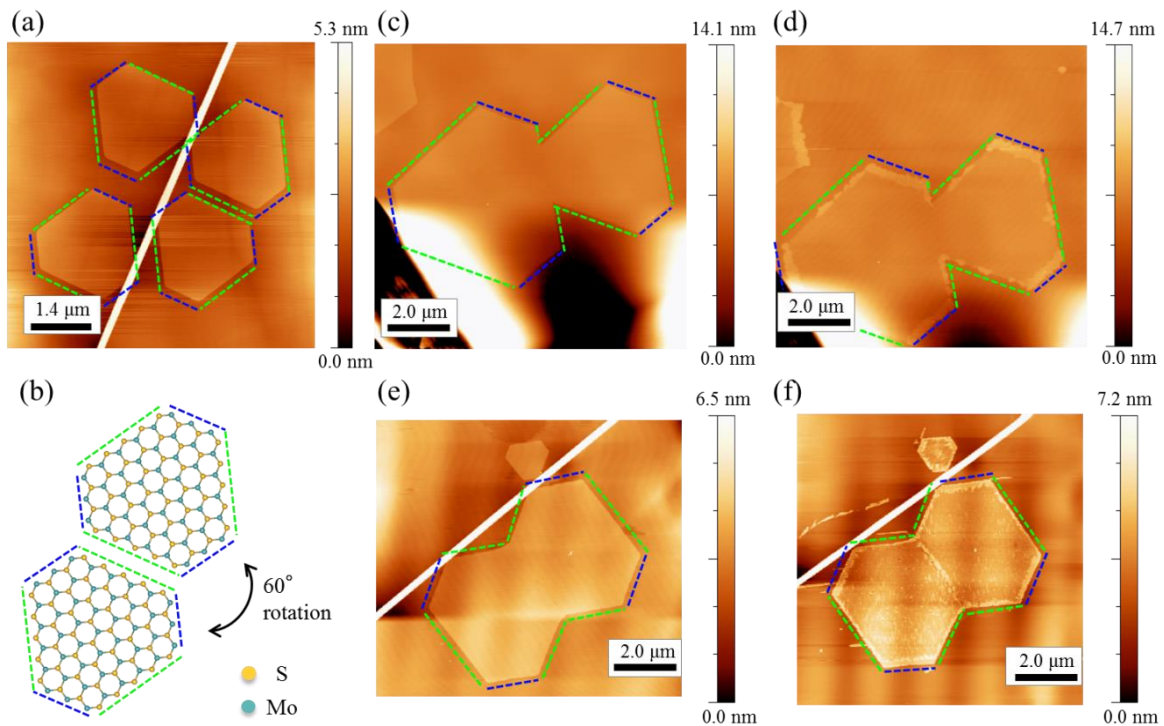

Figure 1. (a) An AFM image of monolayer MoS$_2$ grown on h-BN. Long and short edges of MoS$_2$ flakes are marked by green and blue dotted line, respectively. (b) Structural model of the grown MoS$_2$ with relative angle of 60° degree. (c), (d) An AFM image of MoS$_2$ before and after the oxidation. The relative angle between the two grains in this case is 0°. (e), (f) An AFM image of MoS$_2$ before and after the oxidation. The relative angle between the two grains in this case is 60°. White linear contrasts in Figure 1(a), (c), and (f) are wrinkles in hBN flakes.

To investigate the structure and local electronic structure of GBs, we performed STM/STS measurements around the GBs. For this purpose, we use MoS$_2$ grown on graphite, where the same orientation-limited growth of MoS$_2$ occurs. Figure 2(a) is a STM image of a MoS$_2$ grown on graphite, where positions of GB-0° are highlighted by arrows; we confirmed the monolayer structure by a line profile analysis at the edge (Figure S3). As can be seen, the GB-0° image is slightly darker than its peripheral place, which indicates that GB-0° has different local density of states from its peripheral place. For detailed investigations of structure and electronic structure, atomic-resolution STM observation of the GB-0° was carried out. Figure 2(b) is a STM image of the GB-0° at high magnification, showing the triangular array of sulfur atoms as bright spots. Based on a close inspection of the STM image, the misorientation angle between the two domains is almost zero (Figure S4 and S5). GB-0° is imaged as slightly darker than its peripheral place at the middle of the STM topographic image, and we observed no defects at the GB-0°; neither vacancies nor insertion of atomic rows are seen. The well-stitched structure of GB-0° revealed by STM observation is consistent with its observed low reactivity toward oxidation reactions. It should be noted that a translational mismatch should exist at the boundary even with orientation matching between two grains. This result, however, clearly demonstrates that GB-0° has a stitched structure without defects, indicating that most of the translational mismatch can be relaxed through sliding on the graphite plane. The ultraflat surfaces of graphite and $MoS_2$ may lead to ultralow friction between them, which should facilitate the sliding.[41-43]

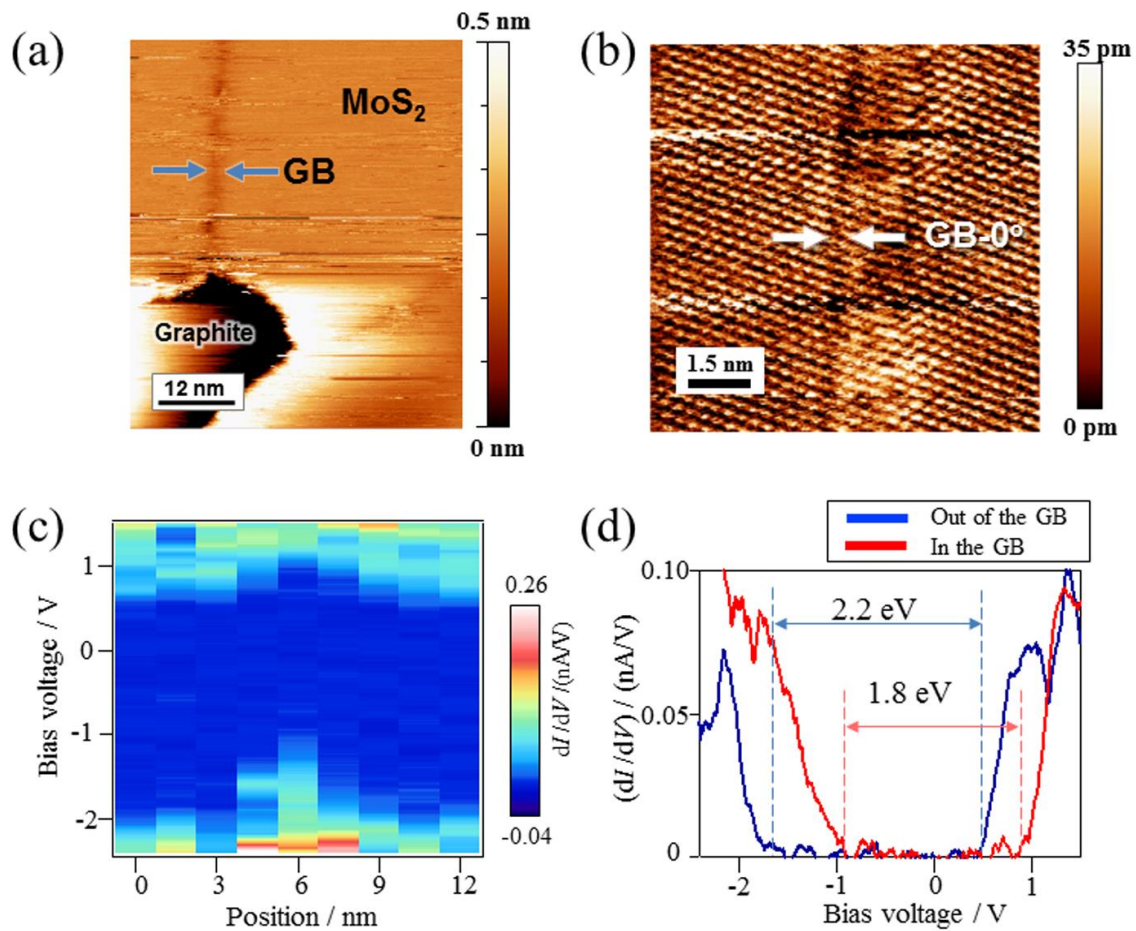

Figure 2. (a) An STM image of MoS$_2$ grown on graphite. The blue arrows indicate the position of GB. (b) A magnified STM image around the GB-0°. The GB-0° can be imaged as a dark linear contrast, which are indicated by the arrow. (c) dI/dV mapping across the GB-0°. (d) STS spectra measured at a place out of the GB and on the GB.

As demonstrated by darker contrasts in the STM image, even though GB-0° has a well-stitched structure, the local electronic structure at GB-0° is different from that of its peripheral places. To see the differences in the electronic structures, we carried out STS and dI/dV mapping to visualize the local density of states. Figure 2(c) shows a dI/dV map across the GB-0°, which is located at a lateral position of around 6 nm in Fig. 2(c). As clearly seen in the figure, both CBM and VBM show upward shifts at the GB-0°. Figure 2(d) is a STS spectrum at GB-0°, showing that the upward shift at VBM (0.8 eV) is larger than that at CBM (0.4 eV). This results in a reduction of the bandgap at the GB-0° from 2.2 eV to 1.8 eV; the observed bandgap of pristine monolayer $MoS_2$ (2.2 eV) is consistent with the value reported previously: 2.15–2.4 eV.[44-46]

Because GB-0° has a stitched structure, this upward shift cannot be explained by formation of defects-mediated midgap states and can probably be explained by the local strain at the GB-0°. The bandgap of monolayer $MoS_2$ is very sensitive to strain, and strain causes bandgap narrowing through upward/downward shift of VBM/CBM[47]. The observed modulation of bands, however, is upward shift in both CB and VB, which probably originates from accumulation of electrons at the boundary. This discrepancy can be understood if piezoelectric charge is taken into account[48]. As monolayer $MoS_2$ has a non-centrosymmetric structure, the local strain can induce charge accumulation at the GB-0°, leading to the observed upward shift of CBM and VBM. One important implication is that a small strain, which probably arises from residual translational mismatch even after the sliding-based relaxation, remains at GB-0°, where the local electronic structure is strongly altered.

To investigate the degree of strain at GB-0°, we performed detailed image analyses with the high-resolution STM image shown in Figure 2(b). Figure S4 shows a contrast-enhanced STM image after applying high-pass filter to filter out the low-frequency noise. It is clear that there are no atomic defects at GB-0°. A line profile along the yellow line clearly demonstrates that location of bright spots in the STM image align periodically without noticeable distortion. In addition, we performed fast Fourier transform (FTT) analysis on the STM image shown in Figure 2(b). As shown in Figure S5, a FFT image at GB-0° shows spots with 6-fold symmetry, which is consistent with a triangular lattice of the sulfur array. The 6-fold symmetric pattern in the FFT image of GB-0° is almost identical to a FFT image at a corresponding peripheral place; line profiles along the green arrows in the FFT image at GB-0° and the peripheral place also coincide well. This means that the difference in lattice constants at the GB-0° and its peripheral place is less than the experimental resolution (2%). As discussed above, the observed difference in bandgap at GB-0° and peripheral places is 0.4 eV. Even though we assume that the difference in bandgap originates only from lattice strain, the strain should be comparable to the experimental resolution, and it is difficult to image the strain directly.[49] These analyses mean that small distortion less than the experimental resolution can remain at the GB-0°, and significant bandgap modulation can occur even in the case of GB-0°. This suggests that it is important to grow large single crystal of TMDs without any boundaries for future application with high-mobility TMD films.

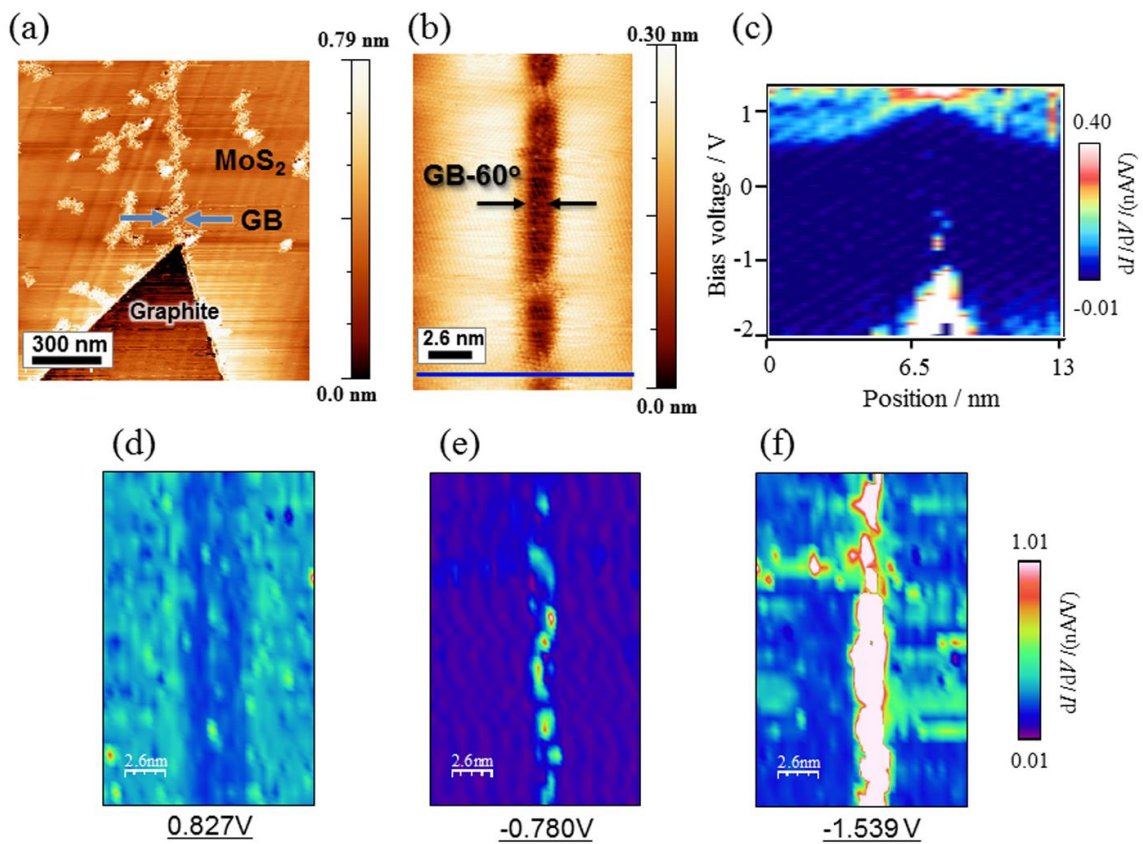

Figure 3. (a) An STM image of $MoS_2$ grown on graphite with a grain boundary of 60°. (b) A magnified STM image around the GB-60°. (c) dI/dV mapping across the GB-60° measured along the line in (b). (d), (e), and (f) STS spectral image of the GB-60° measured with different bias voltage of 0.827, -0.780, and -1.539 V, respectively.

In the case of GB-60°, structural defects exist and the local electronic structure is strongly modified. Figure 3(a) is a STM image of $MoS_2$ on graphite near the GB-60°. The vertical linear contrast at the middle of the image corresponds to an impurity attached at GB-60°, where strong binding sites for impurities should exist. Figure 3(b) is a magnified STM image of clean GB-60°, where the atomic structure can be seen. Based on close investigation of the STM image, we found that the angle between GB-60° and the zigzag edge of $MoS_2$ is about 20°. Figure 3(c) shows the STS spectral mapping along the blue line in Figure 3(b). As clearly seen in Fig. 3(c), the electronic structure is significantly modulated at GB-60°, where both CBM and VBM upshift to reduce the bandgap from 2.3 eV to 1.9 eV. To investigate the spatial distribution of the boundary state at GB-60°, we performed dI/dV mapping at three different bias voltages of 0.83, -0.78, and -1.54 V. Figures 3(d), (e), and (f) show the observed dI/dV mappings at bias voltages of 0.83, -0.78, and -1.54 V, respectively. As clearly seen, the boundary state strongly localizes at GB-60°. In addition, the boundary state corresponding to a bias voltage of -0.78 V shows a dotted distribution rather than a linear uniform distribution, and this means that the boundary state originates from a specific defect site existing at the GB-60°.

## 4. CONCLUSION

In this paper, electronic properties and defect densities in two types of GBs in $MoS_2$ grown by the CVD process were investigated. The orientations of $MoS_2$ grown on hBN and graphite by the CVD process are limited to two directions and the misorientation angles of the two flakes are 0° and 60°. It is confirmed that two grains are stitched completely in the GB-0°, but have an upshift of band structure due to the local stress and charge accumulation. In the GB-60°, the structure of GB is clearly imaged by STM/STS without the disturbance of adsorbates on GB. The band structure in GB-60° upshifts and localized states appear. In the case of two grains of $MoS_2$ stitched at the same angle, the electronic structure of GB-0° is modified due to local stress and carrier accumulation. It will be a challenge to make a $MoS_2$ sheet without modulation of the electronic state or the structure.

Acknowledgement

This work was supported by JSPS KAKENHI Grant numbers JP16H06331, JP16H03825, JP16H00963, JP15K13283, JP25107002, and JST CREST Grant Number JPMJCR16F3. K.W. and T.T. acknowledge support from the Elemental Strategy Initiative conducted by the MEXT, Japan and the CREST (JPMJCR15F3), JST.

Manoharan, and X. L. Zheng  Optoelectronic crystal of artificial atoms in strain-textured molybdenum disulphide *Nat Commun* **2015** **6**

Electronic Supplementary Information

# The Atomic and Electronic structure of 0° and 60° grain boundaries in MoS$_2$

Terunobu Nakanishi[1], Shoji Yoshida[2], Kota Murase[2], Osamu Takeuchi[2], Takashi Taniguchi[3], Kenji Watanabe[3], Hidemi Shigekawa[2], Yu Kobayashi[4], Yasumitsu Miyata[4], Hisanori Shinohara[1], and Ryo Kitaura[1,*]

1 Department of Chemistry, Nagoya University, Nagoya 464-8602, Japan
2 National Institute for Materials Science, 1-1 Namiki, Tsukuba 305-0044, Japan
3 Faculty of Pure and Applied Sciences, University of Tsukuba, Tsukuba 305-8571, Japan
4 Department of Physics, Tokyo Metropolitan University, Hachioji, Tokyo 192-0397, Japan

*Corresponding Author: R. Kitaura
Tel: +81-52-789-2482, Fax: +81-52-747-6442,
E-mail: r.kitaura@nagoya-u.jp

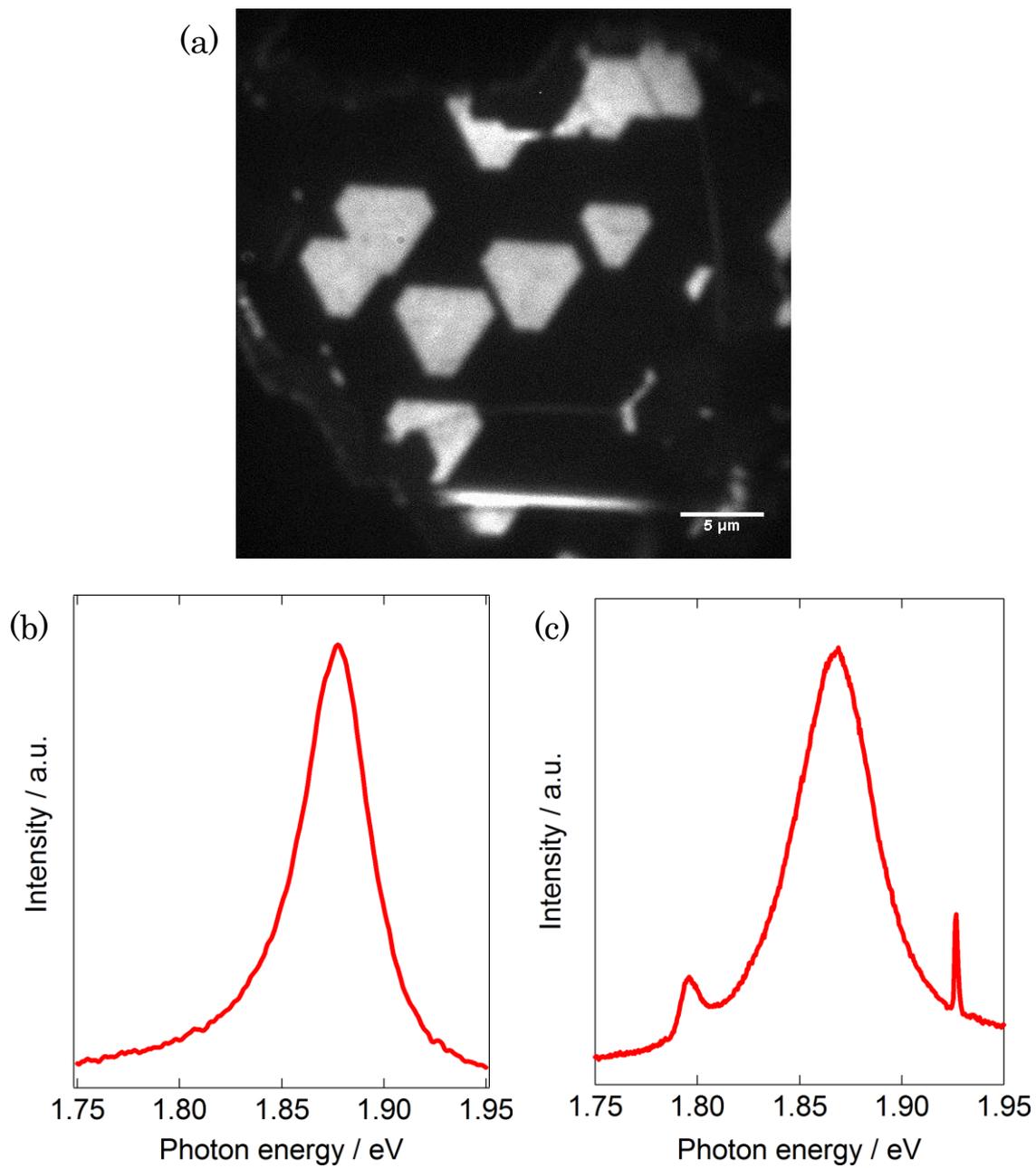

Figure S1. (a) a typical PL image of MoS$_2$ grown on a hBN flake measured at room temperature. (b), (c) a typical PL spectrum of monolayer MoS$_2$ grown on hBN and graphite measured at room temperature.

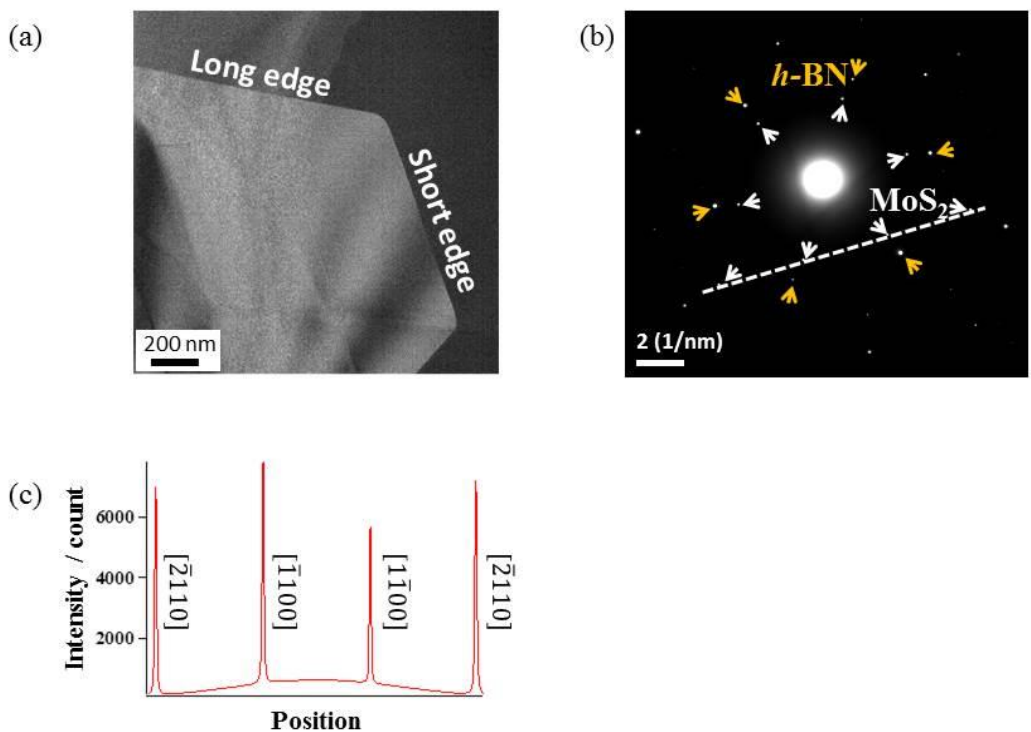

Figure S2. (a) a TEM image of MoS$_2$/hBN, (b)a corresponding electron diffraction image, and (c) an intensity profile of the electron diffraction along the dotted line in (b). The diffraction pattern clearly demonstrates that the crystal orientations of the grown MoS$_2$ flake and the hBN flake underneath perfectly matches. The relation between the TEM image and the diffraction pattern also demonstrates that edges of the MoS$_2$ correspond to the zigzag direction. The intensity profile of the electron diffraction shows intensity difference in the diffraction spots, which leads to assignment of the edge structure of the MoS$_2$; long edges and short edges are S-edges and Mo-edges, respectively.

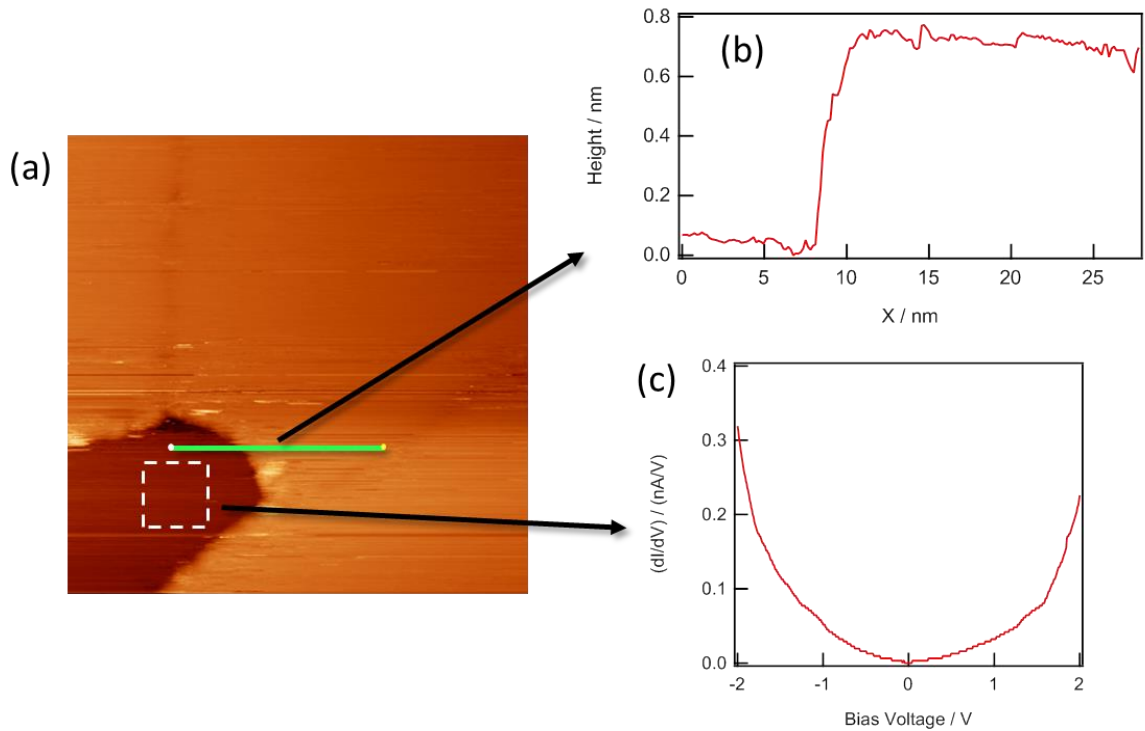

Figure S3. (a) a STM image of $MoS_2$/graphite (this figure is the same as Fig. 2(a)) (b) a line profile along the green line in the Fig. S3(a). (c) a STS spectrum taken at the white dotted square in the Fig. S3(a). As shown in the Fig. S3(b) and S3(c), height of the MoS2 layer from the graphite surface is about 0.7 nm, which is consistent to monolayer structure.

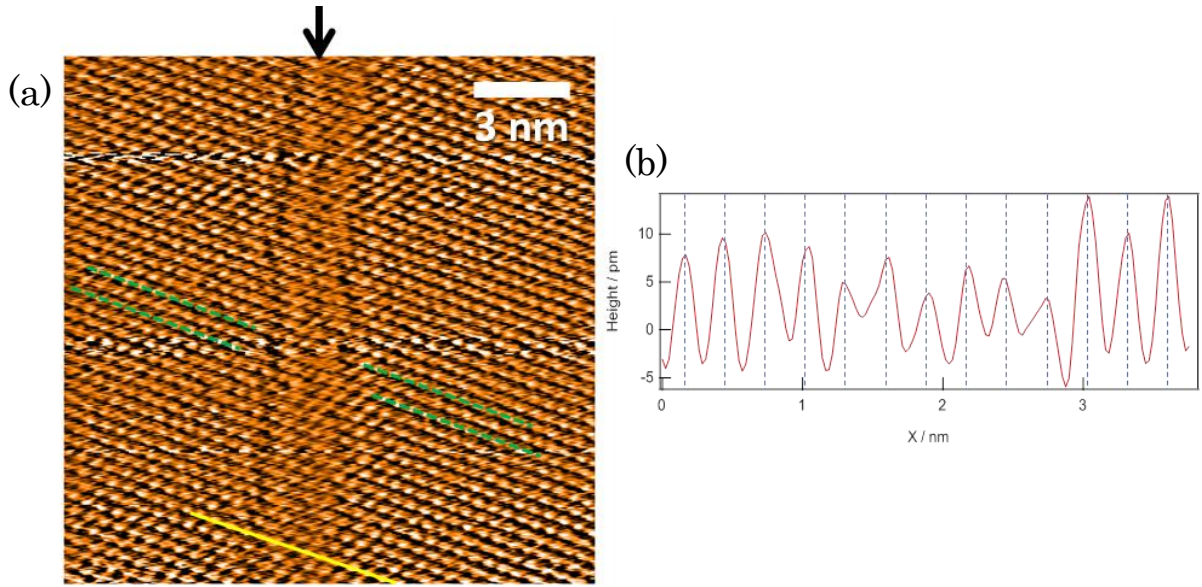

Figure S4. (a) A contrast-enhanced STM image of GB-0° in MoS$_2$/graphite. High-pass filter was applied to filter out the low-frequency noise. (b) a line profile along the yellow line in the Fig. S3(a). The dashed lines in the Fig. S3(a) are drawn along the rows of spots, which arise from S atoms in MoS$_2$. Dashed lines on the left side (the left domain) and the right side (the right domain) are parallel, which clearly demonstrates that the misorientation between two grains is almost zero. In addition, the line profile demonstrates that there is no noticeable change in S-S distances.

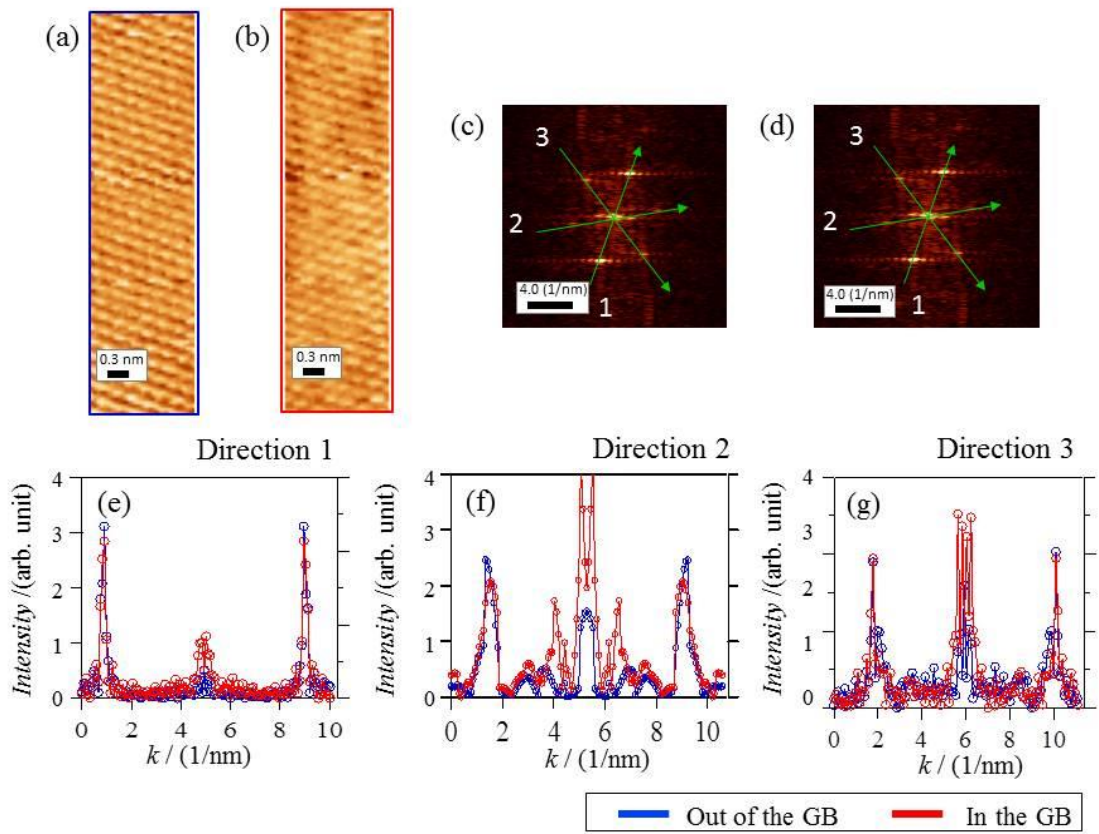

Figure S5. (a) and (b) correspond to STM images of MoS$_2$ close to and at the GB-0°, respectively. (c) and (d) correspond to FFT images of (a) and (b). (e), (f), and (g) are line profile along arrows in (c) and (d). The profiles shown in blue and red are almost identical, and this means that difference in lattice constants in the two regions is less than the experimental resolution.

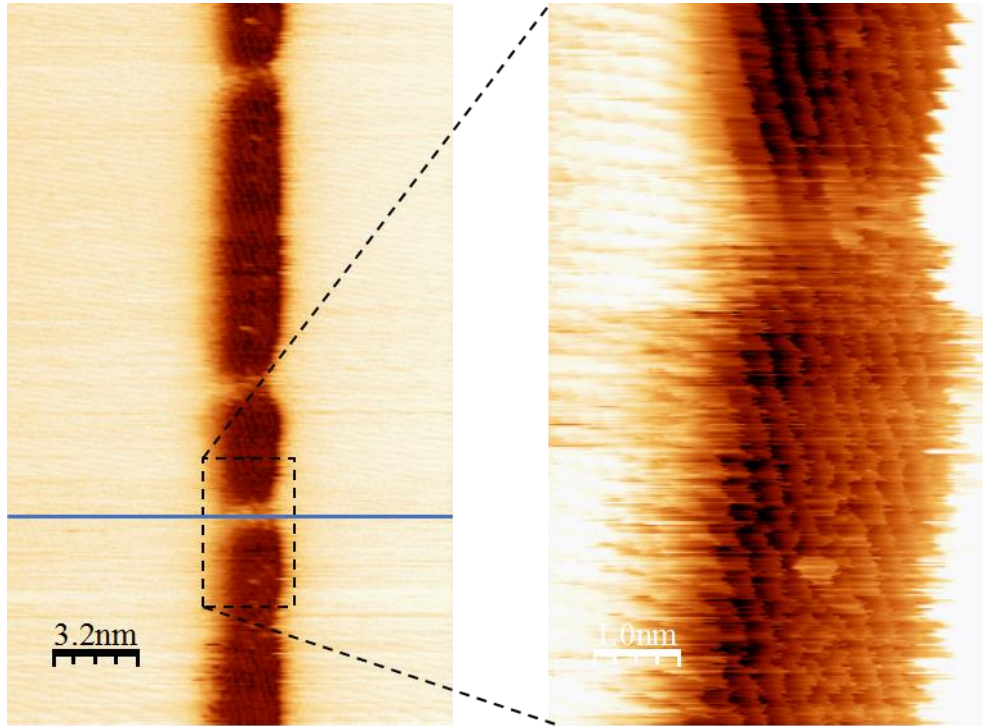

Figure S6. An atomic-resolution STM image around the GB-60°. The blue line corresponds to the place where the STS line profile shown in Fig. 3(c) is measured. It is clear that two domains are connected to form a domain boundary, GB-60°.